\documentclass[a4paper,11pt]{article}
\usepackage[toc,page]{appendix}
\usepackage{graphicx}
\usepackage{colortbl}
\usepackage{amsmath}
\usepackage{subfigure}
\usepackage{mathtools}

\DeclarePairedDelimiter\ket{\lvert}{\rangle}
\DeclarePairedDelimiterX\braket[2]{\langle}{\rangle}{#1 \delimsize\vert #2}
\usepackage[utf8]{inputenc}
\usepackage{float}
\usepackage[normalem]{ulem}
\usepackage{epstopdf}
\usepackage{amsfonts,amsmath,amssymb}
\usepackage{hyperref}

\usepackage{epsfig,multicol,bbm}
\usepackage{color}
\usepackage[dvipsnames]{xcolor}

\definecolor{darkblue}{rgb}{0.0, 0.0, 0.55}
\definecolor{grey}{rgb}{0.57, 0.64, 0.69}
\definecolor{lightbrown}{rgb}{0.71, 0.4, 0.11}

\newcommand{\tcr}{\textcolor{red}}

\def\ndelta{\delta\hspace{-0.50em}\slash\hspace{-0.05em} }

\newcommand{\be}{\begin{equation}}
\newcommand{\ee}{\end{equation}}

\date{}

\newcommand\fverb{\setbox\pippobox=\hbox\bgroup\verb}
\newcommand\fverbit{\egroup\item[\fbox{\unhbox\pippobox}]}

\newbox\pippobox

\begin{document}
\title{Chiral Quartic Massive Gravity in Three Dimensions}
\author{Seyed Naseh Sajadi\thanks{Electronic address: naseh.sajadi@gmail.com}\,,\,
Supakchai Ponglertsakul\thanks{Electronic address: supakchai.p@gmail.com}\,,\,
\\
\small Strong Gravity Group, Department of Physics, Faculty of Science, Silpakorn University,\\ Nakhon Pathom 73000, Thailand\\
}
\maketitle 


\begin{abstract}

We study New Massive Gravity (NMG) with Chern–Simons (CS), cubic, and quartic terms under the Compère–Song–Strominger (CSS) boundary conditions. By employing a semi-product of a Virasoro and a $U(1)$ Kac-Moody current algebra as the asymptotic symmetry algebra, we calculate the entropy of BTZ black holes via the degeneracy of states belonging to a Warped-CFT. Then, we compute the linearized energy excitations using the representations of the algebra $U(1)\times SL(2, R)_{R}$ and demonstrate that the energies of excitations are non-negative at two chiral points in the parameter space. 
\end{abstract}

\maketitle
\section{Introduction}

The absence of a complete theory of quantum gravity has motivated extensive studies of lower-dimensional models as laboratories for exploring its fundamental structure. In this regard, three-dimensional (3D) gravity provides a particularly useful setting. Despite 3D gravity with a cosmological constant lacking local degrees of freedom, it does possess global degrees of freedom, most notably through the existence of the Banados-Teitelboim-Zanelli (BTZ) black hole solutions \cite{BTZ}. The first extension of cosmological 3D gravity is Topologically Massive Gravity (TMG), achieved by supplementing the action with an odd-parity gravitational Chern-Simons term \cite{TMG1, TMG2}. The TMG introduces a single propagating massive graviton mode with definite helicity and is power-counting renormalizable.
Holographically, cosmological TMG (CTMG) admits $2+1$-dimensional anti-de Sitter (AdS) solutions that are dual to a two-dimensional conformal field theory (CFT) with two copies of the Virasoro algebra. Despite these successes, the model faces persistent challenges, nonunitarity introduced by near-boundary logarithmic modes at the chiral point \cite{MaloneyStrominer},\cite{Grumiller:2008qz}.
The chiral limit of CTMG \cite{Chiralgravity} is of particular importance, as it resolves the long-standing bulk–boundary unitarity conflict between positive boundary central charges and the existence of a well-defined bulk massive spin-2 mode. However, in this limit, one of the two boundary central charges vanishes, giving rise to logarithmic excitations that render the dual CFT nonunitary.
As an alternative, New Massive Gravity (NMG) is introduced as a higher-curvature, parity-preserving modification of 3D gravity \cite{Bergshoeff:2009hq}. At a linearized level, the NMG describes a massive graviton with the same dynamics as the Fierz-Pauli theory, but enforcing bulk unitarity inevitably compromises the unitarity of the boundary theory. Further generalizations, collectively referred to as Extended New Massive Gravity (ENMG), have been proposed in \cite{Sinha:2010ai,Gullu:2010pc,Afshar:2014ffa,Yekta:2024aac}, including versions coupled to a Maxwell field in \cite{Paulos:2010ke,Moussa:2008sj}.

It is well understood that boundary conditions play a crucial role in determining the asymptotic symmetry structure of any gravitational theory. Recent investigations extending beyond the standard Brown-Henneaux boundary conditions \cite{BrownHenneaux} have explored various alternative possibilities \cite{BeyonBH1,BeyonBH2,BeyonBH3,BeyonBH4,BeyonBH5,BeyonBH6,BeyonBH7,BeyonBH8,Dengiz:2020fpe,Bertin:2012qw}, leading to the intriguing suggestion that a two-dimensional CFT may not, in fact, serve as the boundary theory of $2+1$-dimensional pure AdS space. In this context, the boundary conditions proposed by Compere, Song, and Strominger (CSS) have attracted particular interest \cite{BeyonBH2}. Specifically, the CSS demonstrated that, under a certain class of alternative boundary conditions, the asymptotic symmetry algebra of a $2+1$ dimensional theory becomes the semidirect product of a Virasoro algebra with a $U(1)$ Kac-Moody algebra, the symmetry structure characteristic of two-dimensional warped conformal field theories (WCFT) \cite{BeyonBH4, Detournay2}.

Ciambelli, Detournay, and Somerhausen have shown that imposing CSS boundary conditions on the TMG leads to two critical points in the space of coupling constants, at which the asymptotic symmetry algebra reduces to a chiral Virasoro algebra or a $U(1)$ Kac-Moody algebra \cite{NewChiralGravity}. Following this approach, the analysis of general minimal massive gravity (GMMG) under the CSS boundary conditions is carried out in \cite{Setare:2021ugr}. In the present work, we extend this investigation to quartic gravity theory within the CSS framework.
We derive the entropy of BTZ black holes by counting the degeneracy of states in the dual Warped-CFT. Furthermore, we compute the linearized energy excitations and show that their energies remain non-negative at two critical points in parameter space, corresponding to cases where the charge algebra reduces to either a Virasoro or a Kac-Moody algebra. Finally, we explore special limits of the quartic theory that reproduce known $2+1$-dimensional massive gravity models.
It should be noted, however, the CSS boundary conditions determine the form of the asymptotic symmetry algebra, and the higher curvature terms in the bulk theory do not introduce new symmetries, but they still have an important role. They modify the covariant phase space charges and the associated central extensions, which leads to shifting in the values of the conserved charges and symmetry coefficients. As a result, the physical quantities such as energy, angular momentum, entropy, as well as unitarity conditions and the holographic data of the dual theory, receive contributions from the bulk theory rather than only from the choice of boundary conditions. Moreover, including higher-curvature terms—with their larger numbers of coupling constants—allows us to solve the tension between the unitarity of the bulk and that of the boundary theories. Therefore, the physics is not determined purely from the boundary conditions, and the bulk dynamics are important.

The structure of the paper is as follows. In Sect. \ref{sectwo}, we analyze the charge algebra of quartic gravity under the CSS boundary conditions and compute the entropy of BTZ black holes by evaluating the degeneracy of states in the dual Warped-CFT. In Sect. \ref{secthree}, we determine the energy spectrum of linearized gravitons in the AdS background. Finally, Sect. \ref{secconc} presents our conclusions and outlines possible directions for future research.

\section{Quartic Theory Under CSS Fall-Offs}\label{sectwo}
The action of NMG, CS, and the cubic and quartic terms is given by
\begin{equation}
    I=\dfrac{1}{2\kappa}\int \sqrt{-g}\left(R-2\bar{\Lambda}+ \mathcal{L}_{CS}+\mathcal{L}_{QUD}+\mathcal{L}_{CUB}+\mathcal{L}_{QUR}\right) d^{3}x, \label{action}
\end{equation}
where $\bar{\Lambda}$ is the bare cosmological constant, $\mu,\eta,\alpha$ and $\beta$ are the coupling constants of CS, NMG, cubic and quartic parts, respectively. The combination of NMG, TMG actions is known as General Massive Gravity (GMG), which has a metric formulation  \cite{Bergshoeff:2009aq}. Each part of the Lagrangian above is given by
\begin{align}
    \mathcal{L}_{CS}=&\dfrac{\mu}{2}\epsilon^{cab}\left(\Gamma^{d}_{c e}\partial_{a}\Gamma^{e}_{d b}+\dfrac{2}{3}\Gamma^{d}_{c e}\Gamma^{e}_{a f}\Gamma^{f}_{d b}\right),\;\;\;\;\;
   \mathcal{L}_{QUD}=\eta_{1} R_{a b}R^{a b}+\eta_{2}R^2,\nonumber\\
  \mathcal{L}_{CUB} =&\alpha_{0}R^3+\alpha_{1}R_{a}{}^{b}R_{b}{}^{c}R_{c}{}^{a}+\alpha_{2} RR_{ab}R^{ab},\nonumber\\
  \mathcal{L}_{QUR}=&\beta_{0}R^4+\beta_{1}R^{2}R^{ab}R_{ab}+\beta_{2}RR^{ab}R_{a}{}^{c}R_{bc}+\beta_{3}R^{ab}R_{a}{}^{c}R_{b}{}^{d}R_{cd}+\beta_{4}\Big(R_{ab}R^{ab}\Big)^2,
\end{align}
and
\begin{align}
\eta_{1}=&\eta,\;\;\;\eta_{2}=-\dfrac{3}{8}\eta,\;\;\;\;\alpha_{0}=-\dfrac{17}{96}\alpha,\;\;\;\alpha_{1}=-\dfrac{2}{3}\alpha,\;\;\;\;\alpha_{2}=\dfrac{3}{4}\alpha,\;\;\;\beta_{1}=-\dfrac{17}{20}\beta-6\beta_{0},\nonumber\\
\beta_{2}=&\dfrac{3}{5}\beta+8\beta_{0},\;\;\;\beta_{3}=-\dfrac{41}{20}\beta-6\beta_{0},\;\;\;\beta_{4}=\dfrac{21}{8}\beta+3\beta_{0}.
\end{align}
This theory is a higher-order curvature deformation of the NMG gravity from the holographic $c-$theorem in the context of AdS/CFT correspondence \cite{Sinha:2010ai}. This action can also be found in the infinitesimal curvature expansion of a Born-Infeld-like action up to the corresponding order \cite{Gullu:2010pc}. By variation of the action to the metric tensor, one can obtain the field equations as follows
\begin{align}
    \mathcal{E}_{\mu\nu}=R_{\mu\nu}-\dfrac{1}{2}g_{\mu\nu}R-\bar{\Lambda}g_{\mu\nu}+\mu\mathcal{C}_{\mu\nu}+\mathcal{E}_{\mu\nu}^{QUD}+\mathcal{E}_{\mu\nu}^{CUB}+\mathcal{E}_{\mu\nu}^{QUR},
\end{align}
where $\mathcal{C}_{\mu\nu}$ is the Cotton tensor
\begin{equation}
    \mathcal{C}_{\mu\nu}=\dfrac{1}{2}\epsilon_{\mu}{}^{\alpha\beta}\nabla_{\alpha}\left(R_{\beta\nu}-\dfrac{1}{4}g_{\beta\nu}R\right),
\end{equation}
and $\mathcal{E}_{\mu\nu}^{QUD}$, $\mathcal{E}_{\mu\nu}^{CUB}$, $\mathcal{E}_{\mu\nu}^{QUR}$ are provided in the appendix \ref{app1}. 
This theory admits more than one maximally symmetric solution for a generic value of the theory's parameter. There exists generally four values of the effective cosmological constant for the solution depending on the parameters $\eta,\alpha,\beta$, and $\bar{\Lambda}$, namely
\begin{equation}
    \bar{\Lambda}-\Lambda-\dfrac{\eta\Lambda^2}{4}-\dfrac{\alpha\Lambda^3}{8}-\beta\Lambda^4=0.
\end{equation}
A unique AdS solution can be obtained by setting the following
\begin{equation}\label{spepo}
    \Lambda=-\dfrac{1}{\ell^2},\;\;\;\bar{\Lambda}=-\dfrac{1}{4\ell^2},\;\;\;\beta=\dfrac{\ell^6}{4},\;\;\;\eta=6\ell^2,\;\;\;\alpha=8\ell^4.
\end{equation}
Our main focus here is to study the theory in the CSS boundary conditions \cite{BeyonBH2},\cite{NewChiralGravity}. For this purpose, let us first recall that the CSS boundary conditions on the metric components are described as  
\begin{align}\label{eqbound}
g_{rr}=&\dfrac{\ell^2}{r^2}+\mathcal{O}\left(\dfrac{1}{r^4}\right),\;\;\;g_{+ -}=-\dfrac{\ell^2 r^2}{2}+\mathcal{O}\left(1\right)\nonumber,\\
g_{r \pm}=&\mathcal{O}\left(\dfrac{1}{r^3}\right),\;\;\; g_{+ +}=\partial_{+}\bar{P}(x^{+})\ell^2 r^2+\mathcal{O}\left(1\right)\nonumber,\\
g_{- -}=&4G\ell \Delta +\mathcal{O}\left(\dfrac{1}{r}\right).
\end{align}
The difference between the CSS and Brown–Henneaux boundary conditions is that CSS is obtained from the Brown–Henneaux conditions by fixing $g_{--}$ and removing the $x^{-}$ dependence of solution space fields. The general solution obeying the CSS boundary conditions can be written as
\begin{align}\label{eqmetric}
ds^2=&\dfrac{\ell^2}{r^2}dr^2-r^2 dx^{+}(dx^{-}-\partial_{+}\bar{P}dx^{+})+4G\ell\left[\bar{L}dx^{+2}+\Delta(dx^{-}-\partial_{+}\bar{P}dx^{+})^2 \right]\nonumber\\
&\;\;\;\;\;\;\;\;\;\;\;\;\;\;\;\;\;-\dfrac{16G^2 \ell^2}{r^2}\Delta \bar{L} dx^{+}(dx^{-}-\partial_{+}\bar{P}dx^{+}),
\end{align}
where $\ell$ stands for the AdS radius, $G$ is the so-called Newton constant, $\bar{L}(x^+)$ and $\partial_+ \bar{P}(x^+)$  are dimensionless periodic chiral boundary functions, and $\Delta$ is an arbitrary constant. Here, $x^{\pm}=\frac{t}{\ell}\pm \phi$ where $\phi \sim \phi +2 \pi $ and the conformal boundary corresponds to the limit as $r \rightarrow \infty$ \cite{BeyonBH2, NewChiralGravity}. One can show that the metric in (\ref{eqmetric}) is a solution to the field equations (corresponding to the action \eqref{action}), provided that 
\begin{equation}
{\bar{\Lambda}=-\dfrac{8\ell^6+2\ell^4\eta+\ell^2\alpha-8\beta}{8\ell^8}.}
\end{equation}
Moreover, by solving the Lie derivative condition along $\xi$ in the allowed region of fall-offs, $\delta g_{\mu\nu}=L_{\xi}g_{\mu\nu}$, one obtains the following asymptotic Killing vectors associated with the metric components in (\ref{eqbound})
\begin{align}\label{killvec}
\xi =&\epsilon \partial_{+}+\left(\sigma +\dfrac{l^2}{2r^2}\partial^{2}_{+}\epsilon \right)\partial_{-}-\bigg(\dfrac{r}{2}\partial_{+}\epsilon\bigg)\partial_{r}+\mathcal{O}\left(\dfrac{l^4}{r^4}\right).
\end{align}
In these expressions, $\sigma(x^{+})$ and $\epsilon(x^{+})$ are arbitrary field-independent chiral functions which are supertranslation in $x^{-}$ direction and $x^{+}$ direction,
respectively. Varying the metric \eqref{eqmetric} along $\xi$, we find the variation of the solution space as follows: 
\begin{align}
    \delta_{\xi}\Delta &= 0\label{eqdel}\\
        \delta_{\xi} \bar{P}^{\prime} &= \left(\epsilon\bar{P}^{\prime}\right)^{\prime}-\sigma^{\prime}\\
        \delta_{\xi} \bar{L} &= \epsilon\bar{L}^{\prime}+2\bar{L}\epsilon^{\prime}-\dfrac{\ell}{4G}\epsilon^{\prime\prime\prime},
\end{align}
where prime denotes derivative with respect to $x^+$.

From \eqref{eqdel}, one can find that $\Delta$ is fixed along the asymptotic symmetry. The second expression suggests that $\bar{P}^{\prime}$ is a $U(1)$ current with level related to the last term. The last expression indicates that $\bar{L}$ is a Virasoro current of weight two, where the last term is the one related to the central extension.

For later convenience, let us introduce a short-hand notation as follows 
\begin{align}\label{symmet}
    \bar{\epsilon}=\epsilon \partial_{+}-\dfrac{r}{2}\partial_{+}\epsilon \partial_{r}+\dfrac{\ell^2}{2r^2}\partial^{2}_{+}\epsilon\partial_{-},\;\;\;\;\; \bar{\sigma}=\sigma\partial_{-}.
\end{align}
At this stage, we have found the variation of the solution space. The surface charges associated with the asymptotic symmetries \eqref{symmet} defined in \cite{Moussa:2008sj,Nam:2010ub,Devecioglu:2010sf,Ghodrati:2016vvf} and the Appendix \ref{surcha} can be computed in the phase space as follows 
\begin{align}
    \ndelta Q_{\bar{\epsilon}} &= \dfrac{1}{2\pi}\int_{0}^{2\pi}d\phi\Bigg[\left(1+\dfrac{\mu}{\ell}+\dfrac{\eta}{2\ell^2}-\dfrac{\alpha}{8\ell^4}+\dfrac{4\beta}{5\ell^6}\right)\delta \bar{L}\nonumber\\
    &~~~~~~~~~~~~~~~~~~~~~~~~-\left(1-\dfrac{\mu}{\ell}+\dfrac{\eta}{2\ell^2}-\dfrac{\alpha}{8\ell^4}+\dfrac{4\beta}{5\ell^6}\right)\left[\Delta\delta(\bar{P}^{\prime 2})+\bar{P}^{\prime}\delta\Delta\right]\Bigg]\epsilon(x^{+}), \\
\ndelta Q_{\bar{\sigma}} &=\dfrac{1}{2\pi}\int_{0}^{2\pi}d\phi\left(1-\dfrac{\mu}{\ell}+\dfrac{\eta}{2\ell^2}-\dfrac{\alpha}{8\ell^4}+\dfrac{4\beta}{5\ell^6}\right)\left[\delta\Delta+\delta\Delta \bar{P}^{\prime}+2\Delta\delta \bar{P}^{\prime}\right]\sigma(x^{+}).
\end{align}
$\ndelta$ indicates that the charges are not integrable in general. To obtain the above surface charges, we first evaluate the surface charges at fixed $(r,x^{+})$ and then at fixed $(r,x^{-})$. Then, the two are combined as $r\to\infty$. These charges are finite but non-integrable. Integrability of the asymptotic charges is required to ensure that the symmetry generators are conserved, and uniquely defined on phase space. Non-integrability of charges generally depends on the particular path that one chooses to integrate on the solution space, not on the final configuration. Here, the non-integrability of the charges originates from the $g_{--}$ component of the fall-offs. Therefore, if $\delta \Delta=0=\delta g_{--}$, the charges become integrable. Moreover, a combination of vectors can be found such that these charges become integrable even when $\delta \Delta \neq 0$ \cite{Aggarwal:2020igb}.
Therefore, in this case $\delta\Delta=0$, the charges read 
\begin{align}
     Q_{\bar{\epsilon}} &= \dfrac{1}{2\pi}\int_{0}^{2\pi}d\phi\Bigg[\left(1+\dfrac{\mu}{\ell}+\dfrac{\eta}{2\ell^2}-\dfrac{\alpha}{8\ell^4}+\dfrac{4\beta}{5\ell^6}\right)\delta \bar{L} \nonumber \\
    &~~~~~~~~~~~~~~~~~~~~~~-
    \left(1-\dfrac{\mu}{\ell}+\dfrac{\eta}{2\ell^2}-\dfrac{\alpha}{8\ell^4}+\dfrac{4\beta}{5\ell^6}\right)\Delta\delta(\bar{P}^{\prime 2})\Bigg]\epsilon(x^{+}),\\
     Q_{\bar{\sigma}}&=\dfrac{1}{\pi}\int_{0}^{2\pi}d\phi \left(1-\dfrac{\mu}{\ell}+\dfrac{\eta}{2\ell^2}-\dfrac{\alpha}{8\ell^4}+\dfrac{4\beta}{5\ell^6}\right)\Delta\delta \bar{P}^{\prime}\sigma(x^{+}).
\end{align}
These charges can now be integrated. We obtain
\begin{align}
Q_{\bar{\epsilon}} &= \dfrac{1}{2\pi}\int_{0}^{2\pi}d\phi\Bigg[\left(1+\dfrac{\mu}{\ell}+\dfrac{\eta}{2\ell^2}-\dfrac{\alpha}{8\ell^4}+\dfrac{4\beta}{5\ell^6}\right)\bar{L} \nonumber \\
&~~~~~~~~~~~~~~~~~~~~~-
\left(1-\dfrac{\mu}{\ell}+\dfrac{\eta}{2\ell^2}-\dfrac{\alpha}{8\ell^4}+\dfrac{4\beta}{5\ell^6}\right)\Delta \bar{P}^{\prime 2}\Bigg]\epsilon(x^{+}), \\
Q_{\bar{\sigma}}&=\dfrac{\Delta}{\pi}\left(1-\dfrac{\mu}{\ell}+\dfrac{\eta}{2\ell^2}-\dfrac{\alpha}{8\ell^4}+\dfrac{4\beta}{5\ell^6}\right)\int_{0}^{2\pi}d\phi  (\bar{P}^{\prime}+C)\sigma(x^{+}). \label{eqq43}
\end{align}
The charge $Q_{\bar{\sigma}}$ has a zero-mode ambiguity and is defined up to a constant $C$ that must be fixed by a choice of background. In the above equation, we fix $C$ by demanding that the charge associated with the exact isometries of (\ref{metricBTZ}) matches with \eqref{eqADMmass}. As a result, we get $C=1/2$. 
For the $U(1)$ sector, the charge algebra is computed as
\begin{equation}
\delta_{\sigma_{2}} Q_{\sigma_{1}}[g]=Q_{[\sigma_{1},\sigma_{2}]}+K_{\sigma_{1},\sigma_{2}}.
\end{equation}
Since $Q_{[\sigma_{1},\sigma_{2}]}=0$, the central extension for the $U(1)$ sector is
\begin{small}
\begin{equation}
K_{\sigma_{1},\sigma_{2}}=-\dfrac{\Delta}{\pi}\left(1-\dfrac{\mu}{\ell}+\dfrac{\eta}{2\ell^2}-\dfrac{\alpha}{8\ell^4}+\dfrac{4\beta}{5\ell^6}\right)\int_{0}^{2\pi} d\phi \sigma_{1}\sigma^{\prime}_{2}.
\end{equation}
\end{small}
Since $\sigma_{1}$ and $\sigma_{2}$ are arbitrary periodic functions, they can be expanded in Fourier modes as $\sigma_{1}=e^{imx^{+}}$, $\sigma_{2}=e^{inx^{+}}$, and calling $ Q_{\underline{\sigma}^{1}}=M_{m} $, $ Q_{\underline{\sigma}^{2}}=M_{n} $, it is easy to obtain
\begin{small}
\begin{equation}
i\left\lbrace M_{m},M_{n} \right\rbrace =m\dfrac{k_{KM}}{2}\delta_{m+n,0},
\end{equation}
\end{small}%
where
\begin{small}
\begin{equation}
 k_{KM}=-4\Delta\left(1-\dfrac{\mu}{\ell}+\dfrac{\eta}{2\ell^2}-\dfrac{\alpha}{8\ell^4}+\dfrac{4\beta}{5\ell^6}\right). \label{keq}
\end{equation}
\end{small}%
This is a centrally extended $U(1)$ algebra with central extension $k$ called the Kac-Moody level. 
For the Virasoro sector, we have
\begin{align}
\left\lbrace Q_{\underline{\epsilon}^{1}},Q_{\underline{\epsilon}^{2}} \right\rbrace &=\delta_{\epsilon_{2}} Q_{\epsilon_{1}}[g], \nonumber \\
&=\dfrac{1}{2\pi}
\int_{0}^{2\pi}d\phi \Bigg[\left(1+\dfrac{\mu}{\ell}+\dfrac{\eta}{2\ell^2}-\dfrac{\alpha}{8\ell^4}+\dfrac{4\beta}{5\ell^6}\right)\bar{L} \nonumber \\
&~~~~~~-
    \left(1-\dfrac{\mu}{\ell}+\dfrac{\eta}{2\ell^2}-\dfrac{\alpha}{8\ell^4}+\dfrac{4\beta}{5\ell^6}\right)\Delta \bar{P}^{\prime 2}\Bigg]\left(\epsilon_{1}\epsilon_{2}^{\prime}-\epsilon_{2}\epsilon_{1}^{\prime}\right)\nonumber\\
    &~~~~~~-\dfrac{\ell}{8\pi G}\left(1+\dfrac{\mu}{\ell}+\dfrac{\eta}{2\ell^2}-\dfrac{\alpha}{8\ell^4}+\dfrac{4\beta}{5\ell^6}\right)\int_{0}^{2\pi}d\phi \epsilon_{1}\epsilon_{2}^{\prime\prime\prime}.
\end{align}
Using the mode decomposition, we take $\epsilon_{1}=e^{imx^{+}}$, $\epsilon_{2}=e^{inx^{+}}$, and calling $Q_{\underline{\epsilon}_{1}}=L_{m}$, $Q_{\underline{\epsilon}_{2}}=L_{n}$, one obtains
\begin{align}
i\left\lbrace L_{m}, L_{n} \right\rbrace =(m-n)L_{m+n}+\dfrac{c_{R}}{12}m^{3}\delta_{m+n,0},
\end{align}%
where
\begin{equation}
    c_{R}=\dfrac{3\ell}{2G}\left(1+\dfrac{\mu}{\ell}+\dfrac{\eta}{2\ell^2}-\dfrac{\alpha}{8\ell^4}+\dfrac{4\beta}{5\ell^6}\right). \label{cReq}
\end{equation}
The generators $M_i$ and $L_i$ satisfy the following classical algebra
\begin{align}\label{algebra}
i\lbrace L_{m},L_{n}\rbrace=&(m-n)L_{m+n}+\dfrac{c_{R}}{12}m^3\delta_{n+m,0},\nonumber\\
i\left\lbrace L_{m},M_{n}\right\rbrace=&-mM_{m+n},\nonumber\\
i\left\lbrace M_{m},M_{n}\right\rbrace=&\dfrac{k_{KM}}{2}m\delta_{n+m,0},
\end{align}

where $k=k_{KM}$ and  $c_R$  are given in \eqref{keq} and \eqref{cReq}, respectively. This algebra is Virasoro–Kac–Moody algebra, which is the symmetry algebra of a WCFT and also is an asymptotic symmetry algebra of 3D gravity with CSS boundary conditions. It should be noted that $c_{R}$ depends only on the couplings of theory, whereas $k_{KM}$, besides couplings of theory, also depends on the geometry $k_{KM}\propto \Delta$. This is because the $U(1)$ current is generated by diffeomorphisms along $\partial_{-}$, and its associated charge is proportional to $g_{--}$, and therefore,  the central term in the charge algebra becomes $k_{KM}\propto \Delta$.
By defining
\begin{equation}\label{muck}
    \dfrac{\mu_{c}}{\ell} \equiv 1+\dfrac{\eta}{2\ell^2}-\dfrac{\alpha}{8\ell^4}+\dfrac{4\beta}{5\ell^6},
\end{equation}
the Kac--Moody level $k_{KM}$ and the right-moving central charge $c_{R}$
can be rewritten as 
\begin{equation}
    k_{KM}=\dfrac{4\Delta}{\ell}\left(\mu-\mu_{c}\right),\;\;\;\;\;\;\;c_{R}=\dfrac{3}{2G}\left(\mu+\mu_{c}\right).
\end{equation}
In addition, when $\mu=\pm\mu_c$, we have
\[
k_{KM}=
\begin{cases}
 0 & \text{if }\;\; \mu=\mu_{c}, \\
 \dfrac{-8\Delta \mu_{c}}{\ell} & \text{if }\;\; \mu=-\mu_{c} ,
\end{cases}
\;\;\;\;\;\;\;\;\;\;\;\;\;
c_{R}=
\begin{cases}
 \dfrac{3\mu_{c}}{G} & \text{if }\;\; \mu=\mu_{c}, \\
 0 & \text{if }\;\; \mu=-\mu_{c} .
\end{cases}
\]
These show, when $\mu=\mu_{c}$, the Kac–Moody sector disappears and the theory becomes purely chiral Virasoro. In this case, the theory behaves like a chiral CFT rather than a WCFT. On the other hand, when $\mu=-\mu_{c}$, the Virasoro sector collapses and the theory is supported purely by a Kac–Moody algebra.

Let us now calculate the entropy of the BTZ black hole by counting the degeneracy of states in the dual 2-dimensional Warped CFT. In this regard, let us first note that the BTZ metric is 
\begin{equation}\label{metricBTZ}
ds^2=-f(r)dt^2+\dfrac{dr^2}{f(r)}+r^2(N(r)dt+d\phi)^2,
\end{equation}
where the metric functions are 
\begin{equation}
f(r)=\dfrac{r^2}{\ell^2}-8GM+\dfrac{16G^2 J^2}{r^2},\;\;\;\; N(r)=-\dfrac{4GJ}{r^2}.
\end{equation}
The black hole horizons are located at the following radii 
\begin{equation}\label{eqqrplmi}
r_{\pm}=\sqrt{2G\ell(\ell M+J)}\pm\sqrt{2G\ell(\ell M-J)}.
\end{equation}
Here $M$ and $J$ are integration constants and for $M>0$, $M\ell>\vert J\vert$ the solution describes a black hole. In quartic theory, the physical mass and angular momentum differ from the integration constants appearing in the metric. The physical mass and angular momentum in the theory, defined as conserved charges at infinity associated with the Killing vectors $\partial_{t}$ and $\partial_{\phi}$, respectively, and are computed using the covariant phase-space method appropriate to the theory (see Appendix \ref{appB}). Equivalently, they can be expressed as the zero modes of the corresponding charge modes as follows:
\begin{align}
    \mathcal{M}=&\dfrac{1}{\ell}(L_{0}+M_{0})=\dfrac{(32\beta-5\alpha\ell^2+20\eta\ell^4+40\ell^6)}{40\ell^8}\left(r_{+}^2+r_{-}^2\right)-\dfrac{2\mu r_{+}r_{-}}{\ell^3}=\dfrac{8G}{\ell^2}\left(\mu_{c}\ell M-\mu J\right), \label{eqADMmass}\\
    \mathcal{J}=&L_{0}-M_{0}=\dfrac{(32\beta-5\alpha\ell^2+20\eta\ell^4+40\ell^6)}{20\ell^7}r_{+}r_{-}-\dfrac{\mu}{\ell^2}(r_{+}^2+r_{-}^2)=\dfrac{8G}{\ell}\left(J\mu_{c}-M\mu\ell\right). \label{eqAngMomentum}
\end{align}
To ensure a positive mass, $\mathcal{M} > 0$, we must have $\mu < \mu_{c}$. The mass and angular momentum of the BTZ black hole at the critical point $\mu=\pm\mu_{c}$ become
\begin{equation}
    \ell \mathcal{M}_{c} = \lvert \mathcal{J}_{c} \rvert ,
\end{equation}
which is precisely the extremality condition for the BTZ black hole.
The Hawking temperature and angular velocity of the black hole are given as
\begin{equation}
    T_{H}=\dfrac{r_{+}^2-r_{-}^2}{2\pi\ell^2 r_{+}},\;\;\;\;\Omega=\dfrac{r_{-}}{\ell r_{+}}.
\end{equation}
The entropy of a black hole is obtained by computing the Noether charge associated with the horizon Killing vector $\xi=\partial_{t}+\Omega\partial_{\phi}$  \cite{Wald:1993nt},\cite{Iyer:1994ys},\cite{Jacobson:1993vj} (see Appendix \ref{appB})
\begin{equation}
S_{BH}=4\pi\left[\left(\ell^6-\dfrac{\alpha\ell^2}{8}+\dfrac{4\beta}{5}+\dfrac{\eta\ell^4}{2}\right)\dfrac{r_{+}}{\ell^6}-\dfrac{\mu r_{-}}{\ell}\right]=\dfrac{4\pi}{\ell}\left(\mu_{c}r_{+}-\mu r_{-}\right). \label{entropy}
\end{equation}
We expect this to be reproduced by counting the number of microstates in the boundary theory.
In the case where $\mu=\mu_{c}$, the BTZ black hole entropy with the Chern-Simons contribution is reproduced \cite{Solodukhin:2005ah,Tachikawa:2006sz,Park:2006gt,Sahoo:2006vz}.
It is straightforward to show that the above thermodynamic quantities satisfy the first law of thermodynamics and the Smarr relation, namely:
\begin{align}
   {d\mathcal{M}=TdS+\Omega d\mathcal{J}},\;\;\;\;\; \mathcal{M}=\dfrac{1}{2}TS+\Omega \mathcal{J}.
\end{align}
These relations ensure the correctness of the thermodynamic quantities. The Cardy formula compatible with CSS falloffs (and asymptotic WAdS$_{3}$) takes the following form
\begin{equation}\label{eqentropy}
{S_{WCFT}=4\pi \sqrt{-M_{0}M_{0vac}}+4\pi \sqrt{-L_{0}L_{0vac}}}
\end{equation}
In this expression, the subscript $vac$ refers to the charges of the vacuum, and here $M = -1/8G$ and $J = 0$ for the global AdS$_{3}$. For the BTZ black hole, one gets the zero modes $(M_0,L_0)$ by solving \eqref{eqADMmass} and \eqref{eqAngMomentum} as follows: 
\begin{align}
L_{0}=&\dfrac{1}{2}\left({\ell \mathcal{M}+\mathcal{J}}\right)=4\left(1-\dfrac{\mu}{\ell}+\dfrac{\eta}{2\ell^2}-\dfrac{\alpha}{8\ell^4}+\dfrac{4\beta}{5\ell^6}\right)\left({\ell {M}+{J}}\right)=4\left(\mu_{c}-\mu\right)\left(M+\dfrac{J}{\ell}\right), \\
M_{0}=&\dfrac{1}{2}\left( \ell \mathcal{M}-\mathcal{J}\right)=4\left(1+\dfrac{\mu}{\ell}+\dfrac{\eta}{2\ell^2}-\dfrac{\alpha}{8\ell^4}+\dfrac{4\beta}{5\ell^6}\right)\left({ \ell {M}-{J}}\right)=4\left(\mu_{c}+\mu\right)\left(M-\dfrac{J}{\ell}\right),
\end{align}
which, for the vacuum, reduces to  
\begin{align}
L_{0vac}=&-\dfrac{\ell}{2}\left(1-\dfrac{\mu}{\ell}+\dfrac{\eta}{2\ell^2}-\dfrac{\alpha}{8\ell^4}+\dfrac{4\beta}{5\ell^6}\right)=\dfrac{1}{2G}\left(\mu-\mu_{c}\right),\\
M_{0vac}=&-\dfrac{\ell}{2}\left(1+\dfrac{\mu}{\ell}+\dfrac{\eta}{2\ell^2}-\dfrac{\alpha}{8\ell^4}+\dfrac{4\beta}{5\ell^6}\right)=-\dfrac{1}{2G}\left(\mu+\mu_{c}\right).
\end{align}
Plugging these into (\ref{eqentropy}) and using \eqref{eqqrplmi}, one finds that $S_{WCFT}$ takes the same form as the black hole entropy $S_{BH}=S_{WCFT}$ \eqref{entropy}. This means, the microscopic entropy counted by the boundary WCFT exactly reproduces the gravitational entropy of the bulk black hole.
This provides strong support for the holographic correspondence under CSS boundary conditions.

{\section{The energy of gravitons}}\label{secthree}

In this section, we are going to study the unitarity of the bulk theory. Therefore, we will obtain the energy of the linearized gravitons in the global AdS$_3$ background. To this end, we  consider the following $2+1$-dimensional AdS spacetime in global coordinates and expressed in light-cone variables 
\begin{equation}\label{eqbackmetric}
ds^{2}=-\dfrac{\ell^2}{4}\left[-4d\rho^2 +dx^{+2}+2\cosh(2\rho)dx^{+}dx^{-}+dx^{-2}\right].
\end{equation}
We define the linearized excitations around the AdS background metric as 
\begin{equation}
g_{\mu \nu}=\bar{g}_{\mu \nu}+h_{\mu \nu},
\end{equation}
where $\bar{g}_{\mu\nu}$ and $h_{\mu\nu}$, respectively, are the background metric (here $AdS_{3}$ metric\footnote{It is important to note that the CSS fall-offs leave the background metric unchanged, and only affect the geometry far away from the boundary.}) and an adequately small perturbation. The linearized equations of motion are as follows
\begin{equation}\label{eqom}
{\cal G}_{\mu \nu}^{(l)}+\bar{\Lambda} h_{\mu \nu}+{\mu}{\cal C}_{\mu \nu}^{(l)}+\eta {\cal E}^{(l)QUD}_{\mu\nu}+\alpha{\cal E}^{(l)CUB}_{\mu\nu}+\beta{\cal E}^{(l)QUR}_{\mu\nu}=0,
\end{equation}
where
\begin{align}\label{eqc}
 {\cal G}^{(l) \mu \nu}=& R^{(l) \mu \nu}-\dfrac{1}{2}g^{\mu \nu} R^{(l)}-2 {\Lambda} h^{\mu \nu},\;\;\;
{\cal C}^{(l) \mu \nu}=\dfrac{1}{\sqrt{-\bar{g}}}\epsilon^{\mu \alpha \beta}\bar{g}_{\beta \sigma}\bar{\nabla}_{\alpha}\left(\mathcal{R}^{(l)\sigma \nu}-\dfrac{1}{4}\bar{g}^{\sigma \nu}\mathcal{R}^{(l)}+2{\Lambda} h^{\sigma \nu}\right)\nonumber\\
\mathcal{R}^{(l)}_{\mu \nu} =&\dfrac{1}{2}\left[-\bar{\nabla}^{2}h_{\mu \nu}-\bar{\nabla}_{\mu}\bar{\nabla}_{\nu}h +\bar{\nabla}_{\mu}\bar{\nabla}_{\sigma}h^{\sigma}_{\nu}+\bar{\nabla}
_{\nu}\bar{\nabla}_{\sigma}h^{\sigma}_{\mu}\right],\;\;\; 
\mathcal{R}^{(l)} =-\bar{\nabla}^{2}h+\bar{\nabla}_{\rho}\bar{\nabla}_{\sigma}h^{\rho \sigma}-2\Lambda h,
\end{align}
where $\bar{\nabla}$ represents the covariant derivative with respect to the background spacetime $\bar{g}_{\mu\nu}$ and $h\equiv h_\mu^\mu$. Therefore, the linearized field equation becomes
\begin{align}\label{eqext}
    \mathcal{E}_{\mu\nu}^{(l)}=&\left(\eta\Lambda^2+\dfrac{\alpha\Lambda^3}{2}+\dfrac{24\beta\Lambda^4}{5}\right)\bar{g}_{\mu\nu}h+\Big(\bar{\Lambda}-\dfrac{11\eta\Lambda^2}{4}-\dfrac{5\alpha\Lambda^3}{4}-\dfrac{57\beta\Lambda^4}{5}\Big)h_{\mu\nu}+\Big(-\dfrac{1}{2}+\dfrac{9\eta\Lambda}{4}+\nonumber\\
    &\dfrac{17\alpha\Lambda^2}{16}+10\beta\Lambda^3\Big)\bar{\nabla}^{2}h_{\mu\nu}-\Big(\dfrac{1}{2}+\dfrac{5\eta\Lambda}{4}+\dfrac{11\alpha\Lambda^2}{16}+\dfrac{34\beta\Lambda^3}{5}\Big)\bar{g}_{\mu\nu}\bar{\nabla}_{\beta}\bar{\nabla}_{\alpha}h^{\alpha\beta}+\Big(\dfrac{1}{2}-\dfrac{\eta\Lambda}{4}-\nonumber\\
    &\dfrac{\alpha\Lambda^2}{16}-\dfrac{2\beta\Lambda^3}{5}\Big)\bar{g}_{\mu\nu}\bar{\nabla}^{2}h-\Big(\dfrac{\eta}{2}+\dfrac{\alpha\Lambda}{4}+\dfrac{12\beta\Lambda^2}{5}\Big)\bar{\nabla}^{4}h_{\mu\nu}-\Big(\dfrac{\eta}{4}+\dfrac{\alpha\Lambda}{8}+\dfrac{6\beta\Lambda^2}{5}\Big)\bar{g}_{\mu\nu}\bar{\nabla}^{2}\bar{\nabla}_{\beta}\bar{\nabla}_{\alpha}h^{\alpha\beta}\nonumber\\
    &+\Big(\dfrac{\eta}{4}+\dfrac{\alpha\Lambda}{8}+\dfrac{6\beta\Lambda^2}{5}\Big)\bar{g}_{\mu\nu}\bar{\nabla}^{4}h+\Big(1+\dfrac{3\eta\Lambda}{2}+\dfrac{7\alpha\Lambda^2}{8}+\dfrac{44\beta\Lambda^3}{5}\Big)\bar{\nabla}_{\mu}\bar{\nabla}_{\alpha}h_{\nu}^{\alpha}+\Big(\eta+\dfrac{\alpha\Lambda}{2}+\nonumber\\
    &\dfrac{24\beta\Lambda^2}{5}\Big)\bar{\nabla}_{\mu}\bar{\nabla}^2\bar{\nabla}_{\alpha}h_{\nu}^{\alpha}-\Big(\dfrac{1}{2}+\dfrac{5\eta\Lambda}{4}+\dfrac{11\alpha\Lambda^2}{16}+\dfrac{34\beta\Lambda^3}{5}\Big)\bar{\nabla}_{\nu}\bar{\nabla}_{\mu}h-\Big(\dfrac{\eta}{4}+\dfrac{\alpha\Lambda}{8}+\dfrac{6\beta\Lambda^2}{5}\Big)\times\nonumber\\
    &(\bar{\nabla}_{\nu}\bar{\nabla}_{\mu}\bar{\nabla}_{\beta}\bar{\nabla}_{\alpha}h^{\alpha\beta}+\bar{\nabla}_{\nu}\bar{\nabla}_{\mu}\bar{\nabla}^2h)+\dfrac{\mu}{2}\Big(-2\Lambda\epsilon_{\mu}{}^{\alpha\beta}\bar{\nabla}_{\beta}h_{\nu\alpha}+\epsilon_{\mu}{}^{\alpha\rho}\bar{\nabla}_{\rho}\bar{\nabla}^2h_{\nu\alpha}-\epsilon_{\mu}{}^{\beta\rho}\bar{\nabla}_{\rho}\bar{\nabla}_{\beta}\bar{\nabla}_{\alpha}h_{\nu}^{\alpha}\nonumber\\&-\epsilon_{\mu}{}^{\beta\rho}\bar{\nabla}_{\nu}\bar{\nabla}_{\rho}\bar{\nabla}_{\alpha}h^{\alpha}_{\beta}+\epsilon_{\mu}{}^{\beta\rho}\bar{\nabla}_{\nu}\bar{\nabla}_{\rho}\bar{\nabla}_{\beta}h\Big)=0,
\end{align}
and at the zero order of perturbation of the field equation, we get
\begin{equation}\label{eqlamef}
    \bar{\Lambda}-\Lambda-\dfrac{\eta\Lambda^2}{4}-\dfrac{\alpha\Lambda^3}{8}-\beta\Lambda^4=0.
\end{equation}
The Chern–Simons term does not contribute to the determination of $\Lambda$, since the Cotton tensor vanishes for any maximally symmetric spacetime.
We fix the gauge freedom by inserting $\bar{\nabla}_{\mu}h^{\mu \nu}=\bar{\nabla}^{\nu}h$ into the linearized field equations, obtaining $h=0$. This gauge is equivalent to the harmonic and traceless gauge  
\begin{equation}\label{eqgau}
\bar{\nabla}_{\mu}h^{\mu \nu}=h=0.
\end{equation}
Upon imposing \eqref{eqlamef} and \eqref{eqgau}, the linearized equation of motion \eqref{eqext} reduces to the following simple form
\begin{align}
&\left({\eta}+\dfrac{\alpha\Lambda}{2}+\dfrac{24\beta\Lambda^2}{5}\right)\bar{\nabla}^{4}h_{\mu\nu}+\left(1-\dfrac{9\eta\Lambda}{2}-\dfrac{17\alpha\Lambda^2}{8}-20\beta\Lambda^3\right)\bar{\nabla}^{2}h_{\mu\nu}\nonumber\\
&+\mu\epsilon_{\mu}{}^{\alpha\beta}\bar{\nabla}^{2}\bar{\nabla}_{\alpha}h_{\beta\nu}-2\mu\Lambda\epsilon_{\mu}{}^{\alpha\beta}\bar{\nabla}_{\alpha}h_{\beta\nu}+\left(-2\Lambda+{5\eta\Lambda^2}+\dfrac{9\alpha\Lambda^3}{4}+\dfrac{104\beta\Lambda^4}{5}\right)h_{\mu\nu}=0.
\end{align}
By introducing the two parameters, 
\begin{align}
m_{1}={\eta}+\dfrac{\alpha\Lambda}{2}+\dfrac{24\beta\Lambda^2}{5},\;\;\;\;\;\;
m_{2}=1-\dfrac{9\eta\Lambda}{2}-\dfrac{17\alpha\Lambda^2}{8}-20\beta\Lambda^3.
\end{align}
 The linearized equations of motion can be rewritten as follows
\begin{equation}\label{eqom1}
\left(\bar{\nabla}^2-2\Lambda\right)\left[\bar{\nabla}^2h_{\mu\nu}+\left(\dfrac{m_{2}}{m_{1}}+2\Lambda\right)h_{\mu\nu}+\dfrac{\mu}{m_{1}}\epsilon_{\mu}{}^{\alpha\beta}\bar{\nabla}_{\alpha}h_{\beta\nu}\right]=0.
\end{equation}
The above linearized equations can also be written in a more compact form as
\begin{equation}
    \left(D^{L}D^{R}D^{M_{1}}D^{M_{2}}h\right)_{\mu\nu}=0,
\end{equation}
with four first-order operators
\begin{align}\label{eqlessmass}
    (D^{L/R})_{\mu}^{\beta}=\delta_{\mu}^{\beta}\mp \dfrac{1}{\sqrt{-\Lambda}}\epsilon_{\mu}{}^{\alpha\beta}\bar{\nabla}_{\alpha},\;\;\;\;\;\;(D^{M_{i}})_{\mu}^{\beta}=\delta_{\mu}^{\beta}+M_{i}\epsilon_{\mu}{}^{\alpha\beta}\nabla_{\alpha},\;\;\;\;(i=1,2).
\end{align}
Thus, solutions are split into four branches. 
First, for massless gravitons, which are also solutions of Einstein gravity ${\cal G}_{\mu \nu}^{(l)}=0$, i.e.,
\begin{align}
    D^{L}D^{R}h_{\mu\nu}=D^{R}D^{L}h_{\mu\nu}=\Big(\bar{\nabla}^2-2\Lambda\Big)h_{\mu\nu}=0.    
\end{align}
Since the linearized Einstein equations in $AdS_{3}$ do not have local bulk degrees of freedom, $D^{L,R}$ corresponds to massless left/right-moving boundary gravitons (massless in the AdS sense). The other two branches are massive gravitons given by
\begin{small}
\begin{align}\label{eqmassive}
           D^{M_{1}}D^{M_{2}}h_{\mu\nu}=D^{M_{2}}D^{M_{1}}h_{\mu\nu}=\bar{\nabla}^2h_{\mu\nu}+\left(\dfrac{1}{M_{1}}+\dfrac{1}{M_{2}}\right)\epsilon_{\beta}{}^{\alpha\mu}\bar{\nabla}_{\alpha}h_{\beta\nu}+\left(\dfrac{1}{M_{1}M_{2}}-3\Lambda\right)h_{\mu\nu}=0.
\end{align}
\end{small}
Then, using \eqref{eqlessmass}, one gets
\begin{equation}\label{eqmasive}
    \bar{\nabla}^{2}h_{\mu\nu}-\left(\dfrac{1}{M_{i}^2}+3\Lambda\right)h_{\mu\nu}=0,\;\;\;\;(i=1,2).
\end{equation}
By comparing \eqref{eqmassive} with \eqref{eqom1}, one obtains
\begin{equation}
    \dfrac{M_{1}+M_{2}}{M_{1}M_{2}}=\dfrac{\mu}{m_{1}},\;\;\;\;\;\;\dfrac{1}{M_{1}M_{2}}=\dfrac{m_{2}}{m_{1}}+5\Lambda
\end{equation}
Solving these equations yields
\begin{equation}
    M_{1}=M_{+}=\dfrac{\mu+\sqrt{\mu^2-20\Lambda m_{1}^2-4 m_{1} m_{2}}}{2(5\Lambda m_{1}+m_{2})},\;\;\;\;\;\;M_{2}=M_{-}=\dfrac{\mu-\sqrt{\mu^2-20\Lambda m_{1}^2-4 m_{1} m_{2}}}{2(5\Lambda m_{1}+m_{2})}.
\end{equation}
These masses $M_{\pm}$ can be associated with gravitons of different helicities and are therefore positive. For the case of $m_{1}=\eta, m_{2}=1$ and $\Lambda=0$, one obtains the masses of the two bulk propagating gravitons in GMG \cite{Bergshoeff:2009hq}.
For $m_{1}=0$ and $m_{2}=1$\footnote{For $\eta=\alpha=\beta=0$, one finds $m_{1}=0$ and $m_{2}=1$, the corresponding theory is topologically massive gravity (TMG).}, we have $M_{+}=\mu$ and $M_{-}=0$, corresponding to TMG, with one massive graviton propagating in the bulk. Moreover, at the chiral point ($\mu=\mu_{c}$), we obtain
\[
M^{c}_{\pm}=
\begin{cases}
 \dfrac{1}{\sqrt{-\Lambda}} \\
 \dfrac{m_{1}\sqrt{-\Lambda}}{m_{2}+5\Lambda m_{1}}
\end{cases}
\;\;\;\;\;\text{OR}\;\;\;\;\;\;\;M^{c}_{\pm}=
\begin{cases}
  \dfrac{m_{1}\sqrt{-\Lambda}}{m_{2}+5\Lambda m_{1}}\\
 \dfrac{1}{\sqrt{-\Lambda}}.
\end{cases}
\]
As can be seen, at these special points, one of the massive modes becomes massless, while the other remains massive. This is consistent with the fact that chirality removes one of the helicities. By now, we have realized that, besides the two massless boundary gravitons, the theory propagates two massive chiral bulk modes whose masses are different. This splitting of masses comes from the presence of the parity-odd gravitational Chern–Simons term in the action \eqref{action}, which breaks parity degeneracy. In the following, we determine the energy associated with each propagating mode.
The fluctuation $h_{\mu\nu}$ can be decomposed as, massive modes $M_i$, left-moving modes $L$ and right-moving modes $R$
\begin{equation}
    h_{\mu\nu}=h_{\mu\nu}^{M_{i}}+h_{\mu\nu}^{L}+h_{\mu\nu}^{R}.
\end{equation}
The quadratic action of $h_{\mu\nu}$, up to total derivative, is
\begin{align}
    S_{2}=&-\dfrac{1}{32\pi}\int d^{3}x\sqrt{-g}h^{\mu\nu}E^{(l)}_{\mu\nu}=\dfrac{1}{64\pi}\int d^3x\sqrt{-g}\Big[-m_{1}\bar{\nabla}_{\lambda}h^{\mu\nu}\bar{\nabla}^{\lambda}\bar{\nabla}^{2}h_{\mu\nu}-{m_{2}}\bar{\nabla}_{\lambda}h^{\mu\nu}\bar{\nabla}^{\lambda}h_{\mu\nu}\nonumber\\
    &-2\Lambda\left({m_{2}}+2\Lambda m_{1}\right)h^{\mu\nu}h_{\mu\nu}-{\mu}\epsilon_{\mu}{}^{\alpha\beta}\bar{\nabla}_{\alpha}h^{\mu\nu}(\bar{\nabla}^2-2\Lambda)h_{\beta\nu}\Big].
\end{align}
The momentum conjugate to $h_{\mu\nu}$ is
\begin{small}
\begin{align}
    \Pi^{(1)\mu\nu}=\dfrac{\delta S_{2}}{\delta(\bar{\nabla_{0}}h_{\mu\nu})}=-\dfrac{\sqrt{-g}}{64\pi}\Big[\bar{\nabla}^{0}\left(2m_{2}h^{\mu\nu}+2m_{1}\bar{\nabla}^{2}h^{\mu\nu}+\mu\epsilon^{\mu\alpha}{}_{\beta}\bar{\nabla}_{\alpha}h^{\beta\nu}\right)
    -\mu\epsilon^{\beta 0 \mu}(\bar{\nabla}^2-2\Lambda)h_{\beta}^{\nu}\Big].
\end{align}
\end{small}
Applying the equation of motion together with \eqref{eqmasive}, we find the momentum conjugate of each mode decomposition
\begin{align}
    \Pi^{(1)\mu\nu}_{M_{i}}=&-\dfrac{\sqrt{-g}}{64\pi}\Big[\Big(\dfrac{2m_{1}}{M_{i}^2}+6\Lambda m_{1}+2m_{2}-\dfrac{\mu}{M_{i}}\Big)\bar{\nabla}^{0}h_{M_{i}}^{\mu\nu}-\mu\Big(\dfrac{1}{M_{i}^2}+\Lambda\Big)\epsilon_{\beta}{}^{0 \mu}h_{{M_i}}^{\beta\nu}\Big)\Big],\\
    \Pi^{(1)\mu\nu}_{L}=&-\dfrac{\sqrt{-g}}{64\pi}\Big[4m_{1}\Lambda+2m_{2}-\mu\sqrt{-\Lambda}\Big]\bar{\nabla}^{0}h^{\mu\nu}_{L},\\
    \Pi^{(1)\mu\nu}_{R}=&-\dfrac{\sqrt{-g}}{64\pi}\Big[4m_{1}\Lambda+2m_{2}+\mu\sqrt{-\Lambda}\Big]\bar{\nabla}^{0}h^{\mu\nu}_{R},
\end{align}
Because we have up to three time derivatives in the Lagrangian \eqref{action}, here we use the Ostrogradsky method\footnote{Since the quartic gravity theory contains time derivatives of higher order, its Hamiltonian formulation cannot be done by performing the standard Legendre transformation on the time derivative of first order. A theory with higher order in time derivatives requires more initial data; hence, more canonical variables in the phase space are required. A well-known approach to treating this is the Ostrogradsky method. The central idea is to take the
several orders in time derivatives of the original coordinate as the new independent coordinates. Then, the Hamiltonian formulation can be obtained by performing the Legendre transformation on all the variables.} \cite{Woodard:2015zca,Bellorin:2025kir}. We are now introducing $K_{\mu\nu}=\bar{\nabla}_{0}h_{\mu\nu}$ as a canonical variable whose conjugate momentum is 
\begin{equation}
\Pi^{(2)\mu\nu}=\dfrac{\delta S_{2}}{\delta(\bar{\nabla}_{0}K_{\mu\nu})}=\dfrac{-\sqrt{-g}g^{00}}{64\pi}\left[-2m_{1}\bar{\nabla}^{2}h^{\mu\nu}+\mu\epsilon_{\beta}{}^{\alpha\mu}\bar{\nabla}_{\alpha}h^{\beta\nu}\right].
\end{equation}
Again, using equations of motion, we have
\begin{align}
    \Pi^{(2)\mu\nu}_{M_{i}}=&\dfrac{\sqrt{-g}g^{00}}{64\pi}\left[\dfrac{2m_{1}}{M_{i}^2}+6m_{1}\Lambda-\dfrac{\mu}{M_{i}}\right]h^{\mu\nu}_{M_{i}},\\
    \Pi^{(2)\mu\nu}_{L}=&\dfrac{\sqrt{-g}g^{00}}{64\pi}\left[4m_{1}\Lambda-\mu\sqrt{-\Lambda}\right]h^{\mu\nu}_{L},\\
    \Pi^{(2)\mu\nu}_{R}=&\dfrac{\sqrt{-g}g^{00}}{64\pi}\left[4m_{1}\Lambda+\mu\sqrt{-\Lambda}\right]h^{\mu\nu}_{R}.
\end{align}
Therefore, the Hamiltonian can be constructed as follows 
\begin{equation}
    \mathcal{H}=\int d^{3}x\left[\dot{h}_{\mu\nu}\Pi^{(1)\mu\nu}+\dot{K}_{\mu\nu}\Pi^{(2)\mu\nu}-S_{2}\right].
\end{equation}
Here, the dot denotes a derivative with respect to time. Finally, using the equations of motion for linearized gravitons, we obtain the generic expressions for the energies as follows:
\begin{align}\label{eqqEnergy}
   E_{M_{i}}&=-\dfrac{1}{32\pi}\left[\dfrac{2m_{1}}{M_{i}^2}+6\Lambda m_{1}-\dfrac{\mu}{M_{i}}+m_{2}\right]\int\sqrt{-g}d^3x\dot{h}_{\mu\nu}^{M_{i}}\bar{\nabla}^{0}h^{\mu\nu}_{M_{i}} \nonumber \\
   &+ \dfrac{\mu}{64\pi}\left[\dfrac{1}{M_{i}^2}+\Lambda\right]\int\sqrt{-g}d^3x\epsilon_{\beta}^{0 \mu}\dot{h}_{\mu\nu}^{M_{i}}h_{M_{i}}^{\beta\nu},\\
    E_{L}&=-\dfrac{1}{32\pi}\left[4m_{1}\Lambda+m_{2}-\mu\sqrt{-\Lambda}\right]\int\sqrt{-g}d^3x\dot{h}_{\mu\nu}^{L}\bar{\nabla}^{0}h^{\mu\nu}_{L},\\
    E_{R}&=-\dfrac{1}{32\pi}\left[4m_{1}\Lambda+m_{2}+\mu\sqrt{-\Lambda}\right]\int\sqrt{-g}d^3x\dot{h}_{\mu\nu}^{R}\bar{\nabla}^{0}h^{\mu\nu}_{R}.
\end{align}
For $m_{1}=0,\;m_{2}=1$, and under the redefinition $\mu\to1/\mu$, the above graviton energy reduces to the expression obtained in \cite{Chiralgravity}.
At the chiral point $\mu=\mu_{c}$, and using $M_{i}=M_{+}^{c}=1/\sqrt{-\Lambda}$ the energies of the massive mode and the left-moving mode vanish, $E_{M_{+}}=E_{L}=0$, while the energy of right-moving mode is given by
\begin{equation}
   E_{R}=-\dfrac{\mu_{c}\sqrt{-\Lambda}}{16\pi}\int\sqrt{-g}d^3x\dot{h}_{\mu\nu}^{R}\bar{\nabla}^{0}h^{\mu\nu}_{R}.
\end{equation}
and the energy of second massive mode becomes 
\begin{equation}
     E_{M_{-}^{c}}=\dfrac{(m_{2}+6m_{1}\Lambda)(m_{2}+4m_{1}\Lambda)}{32\pi m_{1}\Lambda}\left[\int\sqrt{-g}d^3x\dot{h}_{\mu\nu}^{M_{i}}\bar{\nabla}^{0}h^{\mu\nu}_{M_{i}}
   + \dfrac{(m_{2}+4m_{1}\Lambda)}{2m_{1}\sqrt{-\Lambda}}\int\sqrt{-g}d^3x\epsilon_{\beta}^{0\mu}\dot{h}_{\mu\nu}^{M_{i}}h_{M_{i}}^{\beta\nu}\right].
\end{equation}
At the chiral point, this expression remains finite and does not vanish, indicating that the second massive mode continues to propagate in the bulk. In the following, rather than solving the linearized equations of motion directly, we use the symmetry structure of the background to classify the perturbative modes.
The isometry group of AdS$_{3}$, is $SL(2,R)_{L}\times SL(2,R)_{R}$ with generators $\bar{L}_{0,\pm 1}$ and ${L}_{0,\pm 1}$, respectively. We will select a $U(1)\times SL(2, R)_{R}$ sub-algebra compatible with the CSS boundary conditions to classify the perturbation sectors. The $U(1)$ sector is generated by $P_{0}=i\partial_{-}$ and the $SL(2, R)_{R}$ generators is given by
\begin{align}
    L_{0}=i\partial_{+},\;\;\;\;\; L_{\pm 1}=ie^{\pm ix^{+}}\Big[\dfrac{\cosh{2\rho}}{\sin{2\rho}}\partial_{+}-\dfrac{1}{\sin{2\rho}}\partial_{-}\mp\dfrac{i}{2}\partial_{\rho}\Big].
\end{align}
The quadratic Casimir operator of $SL(2, R)$ is $L^2=\frac{1}{2}(L_{1}L_{-1}+L_{-1}L_{1})-L_{0}^2$. When acting on scalars, $L^2+\bar{L}^{2}=\dfrac{1}{2\Lambda}\bar{\nabla}^2$. Thus, using \eqref{eqmasive}, the linearized equation of motion becomes
\begin{equation}
    \left(2\Lambda(L^2+\bar{L}^2)+3\Lambda-\dfrac{1}{M_{i}^2}\right)\left(2\Lambda(L^2+\bar{L}^2)+4\Lambda\right)=0.
\end{equation}
This allows us to use the $U(1)\times SL(2,R)_{R}$ algebra to classify the solutions of \eqref{eqom1}.
The perturbation ansatz compatible with the generators of background along the light-cone directions $x^{+}$ and $x^{-}$ takes the following form
\begin{equation}\label{eqpert}
h_{\mu \nu}(\rho,x^{+},x^{-})=e^{-i(H x^{+}+Px^{-})}f_{\mu \nu}(\rho),
\end{equation}
where the tensor $f_{\mu \nu}$ depends only on the radial coordinate $\rho$, and $H$ and $P$ are eigenvalues associated with translations along $x^{+}$ and $x^{-}$ as
\begin{equation}
{L}_{0}\vert h_{\mu \nu}\rangle = H \vert h_{\mu \nu}\rangle , \;\;\;\;\; {P}_{0}\vert h_{\mu \nu}\rangle = P \vert h_{\mu \nu}\rangle.
\end{equation}
Different values of $H$ and $P$ correspond to different perturbation modes, such as massless, massive, and logarithmic modes.
Using $L^2\ket{h_{\mu\nu}}=-H(H-1)\ket{h_{\mu\nu}}$ for the primary weights, the $(H,P)$ obey
\begin{align}
   & \left(-2(H(H-1)+P(P-1))+3-\dfrac{1}{M_{i}^2\Lambda}\right)\left(-2(H(H-1)+P(P-1))+4\right)=0,\\
   &H-P=\pm 2.
\end{align}
There are two branches of solutions. The first branch has $H(H-1)+P(P-1)-2=0$, which gives two families of solutions as
\begin{align}
    H=\dfrac{3\pm 1}{2},\;\;\;\;\;P=\dfrac{-1\pm 1}{2},\;\;\;\;\; P=\dfrac{3\pm 1}{2},\;\;\;\;\;H=\dfrac{-1\pm 1}{2}.
\end{align}
The solutions with the lower sign diverge at infinity and violate the CSS boundary conditions (see Appendix \ref{app3}). Thus, we will only keep the upper ones which correspond to weights $(2,0)$ and $(0,2)$ associated to left and right moving massless gravitons. This branch corresponds to Einstein-sector modes. The second branch has $-2(H(H-1)+P(P-1))+3-\frac{1}{M_{i}^2\Lambda}=0$ which gives:
\begin{align}
    H=&\dfrac{3}{2}\mp \dfrac{1}{2M_{i}\sqrt{-\Lambda}},\;\;\;\;\;P=-\dfrac{1}{2}\mp\dfrac{1}{2\sqrt{-\Lambda}M_{i}},\label{eqmasswieght1}\\
    P=&\dfrac{3}{2}\mp \dfrac{1}{2M_{i}\sqrt{-\Lambda}},\;\;\;\;\;H=-\dfrac{1}{2}\mp\dfrac{1}{2\sqrt{-\Lambda}M_{i}}.\label{eqmasswieght2}
\end{align}
The only solutions that remain finite at infinity are the plus sign.
For the first massive mode $M_{+}^{c}=1/\sqrt{-\Lambda}$ at the chiral point ($\mu=\mu_{c}$), we obtain $(2,0)$ and $(0,2)$. These are the conformal weights of massless boundary gravitons. For the second massive mode at the chiral point $M_{2}^{c}$ we have
\[
(H^{c}, P^{c})=
\begin{cases}
 \Big(-1-\dfrac{m_{2}}{2m_{1}\Lambda},-3-\dfrac{m_{2}}{2m_{1}\Lambda}\Big) \\
\Big(-3-\dfrac{m_{2}}{2m_{1}\Lambda},-1-\dfrac{m_{2}}{2m_{1}\Lambda}\Big)
\end{cases}
\;\;\;\text{for}\;\;\;\;\;M_{-}^{c}=\dfrac{m_{1}\sqrt{-\Lambda}}{m_{2}+5m_{1}\Lambda}.
\]
Therefore, the second massive graviton remains massive at the chiral point. Using the transverse, traceless, and highest-weight condition, one can obtain $f_{\mu \nu}$, whose components depend on $H, P$, and the integration constants $\mathcal{C}_{1}$ and $\mathcal{C}_{2}$ provided in the Appendix \ref{app3}.
Then, by inserting equations (\ref{eqpert}) with (\ref{eqpert1}) into (\ref{eqqEnergy}) and demanding that the energy $E_i$ should be finite. One ultimately gets:
\begin{itemize}
    \item {Massive mode:} For the massive bulk graviton mode $H=\frac{1}{2}(3+\frac{1}{\sqrt{-\Lambda}M_{i}}),P=-\frac{1}{2}(1-\frac{1}{\sqrt{-\Lambda}M_{i}})$ and $\mathcal{C}_{2}=0$, the energy becomes
\begin{equation}
    E_{M_{MG}}=\dfrac{\Lambda^2\mathcal{C}_{1}^2(M_{i}\sqrt{-\Lambda}+1)(\mu\Lambda M_{i}^3+2M_{i}^2m_{2}+12\Lambda m_{1}M_{i}^2-\mu M_{i}+4m_{1})}{512 G M_{i}^3(2M_{i}\sqrt{-\Lambda}+1)}.
    \end{equation}
    This mode propagates in the bulk and carries a mass $M_{MG}$.
For $m_{1}=0,\;m_{2}=1,\;M_{i}=\mu$ and $\mu\to1/\mu$ the energy of the massive graviton ($E_{M_{MG}}$) corresponds to that provided in \cite{NewChiralGravity}.
\item Right graviton mode: For the right-moving massless graviton $H=2,P=0$ and $\mathcal{C}_{2}=0$, the energy is
\begin{equation}\label{energyRG}
E_{RG}=\dfrac{\mathcal{C}_{1}^2(-\Lambda)^{\frac{5}{2}}}{384G }\left[4m_{1}\Lambda+m_{2}+\mu\sqrt{-\Lambda}\right]=-\dfrac{\mathcal{C}_{1}^2\Lambda^3}{384G}\left(\mu+\mu_{c}\right).
\end{equation}
This graviton is a boundary mode, not a propagating bulk massive graviton. Therefore, its energy entirely depends on the couplings of the theory. At the right chiral point $\mu=-\mu_{c}$, the energy vanishes.
In the limit of $m_{1}=0,\;m_{2}=1$, the energy coincides with the energy of the graviton in TMG provided in \cite{NewChiralGravity}.
\end{itemize}
Finally, we investigate the unitarity of the theory. To check the unitarity of both the bulk and boundary theories, we need to impose
\begin{align}
    c_{R}\geq 0,\;\;\;k_{KM}\geq 0,\;\;\;M_{+}\geq M_{-}\geq 0,\;\;\;E_{M_{MG}}\geq 0,\;\;\;E_{RG}\geq 0.
\end{align}
By rewriting the energy of a massive graviton, one can get:
\begin{equation}
    E_{MG}(M_{\pm})=\dfrac{\Lambda^2\mathcal{C}_{1}^2}{256G}\dfrac{(\mu_{c}\sqrt{-\Lambda}+2m_{1}\Lambda)M_{\pm}-\mu}{M_{\pm}^2(2\sqrt{-\Lambda}M_{\pm}+1)}
\end{equation}
since $M_{\pm}>0$ (no tachyons) and prefactor $\Lambda^2\mathcal{C}_{1}^2/256G>0$, the sign of $E_{MG}$ comes from the second expression
\begin{equation}
    \text{sign}(E_{MG})=\text{sign}\left(\dfrac{(\mu_{c}\sqrt{-\Lambda}+2m_{1}\Lambda)M_{\pm}-\mu}{M_{\pm}^2(2\sqrt{-\Lambda}M_{\pm}+1)}\right)
\end{equation}

\begin{figure}[H]
\centering
\subfigure{\includegraphics[width=0.6\columnwidth]{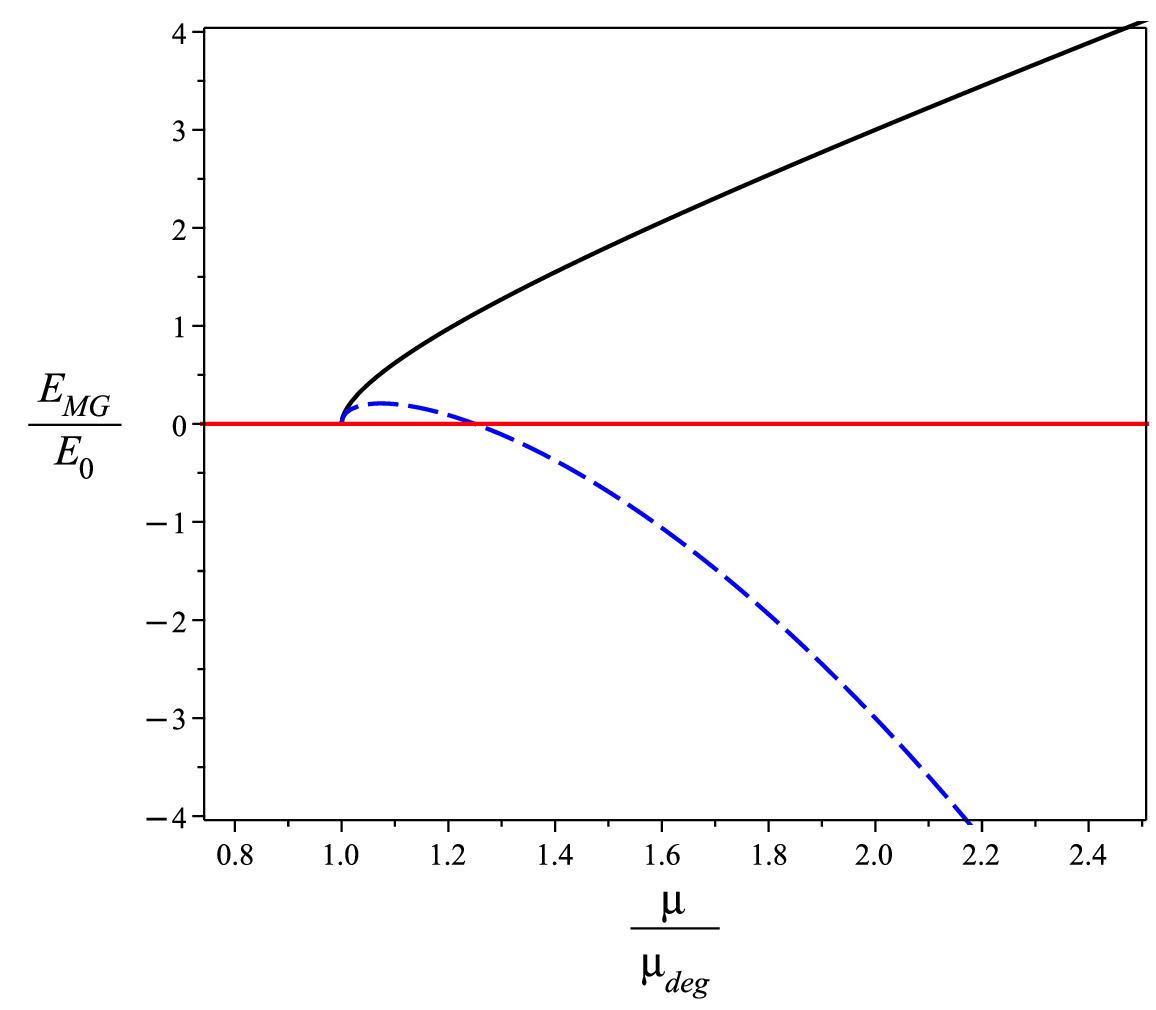}}
\caption{Energy of the massive modes $E(M_{\pm})$ as a function of $\mu/\mu_{c}$. The black solid line corresponds to $M_{+}$, while the dashed blue line corresponds to $M_{-}$.} 
\label{f11kkplot}
\end{figure}

 For $M_{-}$, when $\mu>\mu_{c}$ the energy becomes negative, whereas for $M_{+}$ the energy remains positive. At the chiral point $\mu=\mu_{c}$, the energy of $M_{-}$ vanishes, while the energy of $M_{+}$ is positive.
It is also worthwhile to rewrite $E_{MG}$ in terms of $x=\mu/\mu_{\mathrm{deg}}$ in a simplified form
\begin{equation}
    E_{M_{MG}}(M_{\pm})=E_{0}\sqrt{x^2-1}\left(x\pm\sqrt{x^2-1}-2(x-1)\right),\;\;E
    _{0}=\dfrac{\Lambda^2\mathcal{C}_{1}^2}{1024G}\dfrac{M_{\pm}\sqrt{-\Lambda}+1}{(M_{\pm})^3(2M_{\pm}\sqrt{-\Lambda}+1)}.
\end{equation}
Here, at $\mu=\mu_{\mathrm{deg}}$, we have $M_{+}=M_{-}$. The behavior of the dimensionless graviton energy $E_{MG}/E_{0}$ as a function of the dimensionless ratio $\mu/\mu_{\mathrm{deg}}$ is shown in Fig.~\ref{f11kkplot}.

For $1<x<5/4$, the energies $E_{M_{\pm}}$ are positive. For $x>5/4$, the energy $E_{M_{-}}$ becomes negative, while $E_{M_{+}}$ remains positive.\\
The energy of the right-moving graviton, given in \eqref{energyRG}, is positive for all values of $\mu$; therefore, this mode is always unitary. Finally, for the bulk, we have:

\[
E=
\begin{cases}
 +\;\;\;\; \;\;\text{if}\;\;\;\;\;\;\mu_{deg}<\mu<\mu_{c}\\
0 \;\;\;\; \;\;\text{if}\;\;\;\;\;\;\mu=\mu_{c}\\
-\;\;\;\; \;\;\text{if}\;\;\;\;\;\;\mu>\mu_{c}
\end{cases}
\]

From the boundary side, $c_{R}$ is always positive. Assuming $\Delta>0$, the $U(1)$ Kac–Moody level ($K_{KM}$) is positive for $\mu>\mu_{c}$, vanishes at $\mu=\mu_{c}$, and becomes negative for $\mu<\mu_{c}$.\\

Overall, requiring the positivity of the $U(1)$ Kac–Moody sector and the right-moving central charge implies $\mu\geq\mu_{c}$, whereas the positivity of the bulk energy requires $\mu\leq \mu_{c}$. The only point at which the bulk and boundary conditions are simultaneously satisfied is the chiral point $\mu=\mu_{c}$, where the $U(1)$ sector becomes trivial, and one of the bulk massive modes becomes massless.

\section{Conclusion}\label{secconc}

In this work, we have studied a quartic extension of 3D massive gravity under the CSS boundary conditions, with the goal of addressing tension between bulk and boundary unitarity. By adding parity-even cubic and quartic curvature terms into GMG, we have shown that the bulk higher-curvature couplings lead to nontrivial modifications to the conserved charges, the entropy, and the energy excitations. We identified two chiral points in the parameter space of the theory that correspond to the Virasoro and $U(1)$ Kac-Moody asymptotic symmetry algebra. From the holographic view, we derived the BTZ black hole entropy via state counting in a dual warped conformal field theory using Cardy's formula, showing agreement with the gravitational result using Wald's formula. Then, we analyzed the linearized field equations around the AdS$_3$ background and showed that the theory admits two massless boundary gravitons and two massive bulk gravitons with different helicities. Then, we obtained the generic expressions for the energies of the propagating modes of the theory. We showed that at the chiral point, the $U(1)$ sector becomes trivial, one massive graviton becomes massless, and both bulk and boundary unitarity conditions are satisfied. These results show that higher-curvature bulk terms are necessary to achieve unitarity in three dimensions.

\section*{Acknowledgments}
This research has received funding support from the NSRF via the Program Management Unit for Human Resource and Institutional Development, Research and Innovation grant number $B13F680083$.

\appendix
\section{Field Equation}\label{app1}

By variation of the action with respect to the metric tensor, one can obtain the corresponding equation of motion as follows
\begin{align}
\mathcal{E}_{ab}=\dfrac{1}{\sqrt{-g}}\dfrac{\delta(\sqrt{-g}\mathcal{L})}{\delta g^{ab}}=\mathcal{P}_{acde}\mathcal{R}_{b}{}^{cde}-\dfrac{1}{2}g_{ab}\mathcal{L}-2\nabla^{c}\nabla^{d}\mathcal{P}_{acdb}, \label{LHS}
\end{align}
where $\mathcal{P}^{abcd}=\dfrac{\partial\mathcal{L}}{\partial\mathcal{R}_{abcd}}$. Here, we provide $\mathcal{P}^{abcd}$ for the quadratic, cubic, and quartic parts as shown below.  

For the quadratic and the cubic parts, we have 
\begin{align}
\mathcal{P}^{\alpha\beta\eta\delta}_{\eta_{1}}=&\dfrac{1}{2}\left(g^{\alpha\eta}g^{\beta\delta}-g^{\alpha\delta}g^{\beta\eta}\right)f^{\prime}(R),\\
\mathcal{P}^{\alpha\beta\eta\delta}_{\eta_{2}}=&-\dfrac{1}{2}g^{\beta\eta}R^{\alpha\delta}+\dfrac{1}{2}g^{\beta\delta}R^{\alpha\eta}+\dfrac{1}{2}g^{\alpha\eta}R^{\beta\delta}-\dfrac{1}{2}g^{\alpha\delta}R^{\beta\eta},\\
    \mathcal{P}^{\alpha\beta\eta\delta}_{\alpha_{1}}=&-\dfrac{3}{4}\Big(g^{\beta\eta}R^{\alpha\mu}R^{\delta}_{\mu}-g^{\alpha\eta}R^{\beta\mu}R^{\delta}_{\mu}-g^{\beta\delta}R^{\alpha\mu}R^{\eta}_{\mu}+g^{\alpha\delta}R^{\beta\mu}R^{\eta}_{\mu}\Big),\\
    \mathcal{P}^{\alpha\beta\eta\delta}_{\alpha_{2}}=&\dfrac{1}{2}g^{\alpha\eta}g^{\beta\delta}R_{\mu\nu}R^{\mu\nu}-\dfrac{1}{2}g^{\alpha\delta}g^{\beta\eta}R_{\mu\nu}R^{\mu\nu}-\dfrac{1}{2}g^{\beta\eta}g^{\alpha\delta}R+\dfrac{1}{2}g^{\beta\delta}R^{\alpha\eta}R+\dfrac{1}{2}g^{\alpha\eta}R^{\beta\delta}R\nonumber\\
    &-\dfrac{1}{2}g^{\alpha\delta}R^{\beta\eta}R,
\end{align}

and for the quartic part
\begin{align}
 \mathcal{P}^{\alpha\beta\eta\delta}_{\beta_{1}}=&g^{\alpha\eta}g^{\beta\delta}R_{\mu\nu}R^{\mu\nu}R-g^{\alpha\delta}g^{\beta\eta}R_{\mu\nu}R^{\mu\nu}R-\dfrac{1}{2}g^{\beta\eta}R^{\alpha\delta}R^{2}+\dfrac{1}{2}g^{\beta\delta}R^{\alpha\eta}R^{2}+\dfrac{1}{2}g^{\alpha\eta}R^{\beta\delta}R^2\nonumber\\
 &-\dfrac{1}{2}g^{\alpha\delta}R^{\beta\eta}R^{2},\\
 \mathcal{P}^{\alpha\beta\eta\delta}_{\beta_{2}}=&\dfrac{1}{2}g^{\alpha\eta}g^{\beta\delta}R_{\mu}^{\rho}R^{\mu\nu}R_{\nu\rho}-\dfrac{1}{2}g^{\alpha\delta}g^{\beta\eta}R_{\mu}^{\rho}R^{\mu\nu}R_{\nu\rho}-\dfrac{3}{4}g^{\beta\eta}R^{\alpha\mu}R^{\delta}_{\mu}R+\dfrac{3}{4}g^{\alpha\eta}R^{\beta\mu}R^{\delta}_{\mu}R\nonumber\\
 &+\dfrac{3}{4}g^{\beta\delta}R^{\alpha\mu}R^{\eta}_{\mu}R-\dfrac{3}{4}g^{\alpha\delta}R^{\beta\mu}R^{\eta}_{\mu}R,\\
 \mathcal{P}^{\alpha\beta\eta\delta}_{\beta_{3}}=&-g^{\beta\eta}R_{\mu\nu}R^{\alpha\mu}R^{\delta\nu}+g^{\alpha\eta}R_{\mu\nu}R^{\beta\mu}R^{\delta\nu}+g^{\beta\delta}R_{\mu\nu}R^{\alpha\mu}R^{\eta\nu}-g^{\alpha\delta}R_{\mu\nu}R^{\beta\mu}R^{\eta\nu},\\
 \mathcal{P}^{\alpha\beta\eta\delta}_{\beta_{4}}=&-g^{\beta \eta}R^{\alpha\delta}R_{\mu\nu}R^{\mu\nu}+g^{\beta\delta}R^{\alpha\eta}R_{\mu\nu}R^{\mu\nu}+g^{\alpha\eta}R^{\beta\delta}R_{\mu\nu}R^{\mu\nu}-g^{\alpha\delta}R^{\beta\eta}R_{\mu\nu}R^{\mu\nu}.
\end{align}

\section{Covariant phase space}\label{appB}
Here, we apply the covariant phase-space method to higher-curvature gravity theories, with a particular focus on cubic and quartic terms built from the Riemann and Ricci tensors \cite{Wald:1993nt,Iyer:1994ys,Ashtekar:1987hia,Lee:1990nz,Barnich:2001jy}. The variation of the action to the fields $\phi$ is given by
\begin{equation}
\delta L[\phi]=E_{\phi}\delta\phi+d\Theta(\phi,\delta \phi),
\end{equation}
where $\delta\phi$ is a generic field perturbation and $E_{\phi}=0$ denotes the field equations for the field $\phi$, and $\Theta$ is the symplectic potential picked up from the surface term of the variation. The Noether-Wald charge density, defined by the relation
 \begin{equation}
dQ_{\xi}=\Theta-\xi .L.
\end{equation}
The $(d-2)-$form $k_{\xi}$ can be shown to be explicitly stated
\begin{equation}\label{eqkxi}
k_{\xi}=\delta Q_{\xi}-\xi.\Theta.
\end{equation}
In the next section, we apply this method to cubic and quartic gravity.

\paragraph{Symplectic potential:}
By variation of Lagrangian and using the EoM, the surface $(d-1)$-form $\Theta$ can be read as

\begin{align}
\Theta^{m}_{\alpha_{0}}=&f^{\prime}\left(\nabla_{a}h^{a m}-\nabla^{m}h\right)-\nabla_{a}f^{\prime}h^{m a}+
\nabla^{m}f^{\prime}h,\\
\Theta^{m}_{\alpha_{1}}=&3\nabla_{c}h^{m}_{b}R^{a b}R_{a}^{c}-3h_{b}^{d}\nabla_{d}\left(R^{ab}R_{a}^{m}\right)-\dfrac{3}{2}\nabla^{m}h_{bc}R^{ab}R_{a}^{c}+\dfrac{3}{2}h_{bc}\nabla^{m}\left(R^{ab}R_{a}^{c}\right)-\dfrac{3}{2}\nabla_{b}hR_{a}^{m}R^{ab}\nonumber\\
&+\dfrac{3}{2}h\nabla_{c}\left(R_{a}^{c}R^{am}\right),\\
\Theta^{m}_{\alpha_{2}}=&-\nabla_{a}hRR^{am}+h\nabla_{b}\left(RR^{m b}\right)+2\nabla_{b}h_{a}^{m}RR^{ab}-2h_{a}^{c}\nabla_{c}\left(RR^{am}\right)-
\nabla^{m}h_{ab}RR^{ab} \nonumber\\
&+h_{ab}\nabla^{m}\left(RR^{ab}\right)+\nabla_{c}h^{cm}
R_{ab}R^{ab}-h^{m d}\nabla_{d}\left(R_{ab}R^{ab}\right)-\nabla^{m}hR_{ab}R^{ab}+
h\nabla^{m}\left(R_{ab}R^{ab}\right), \\
\Theta^{m}_{\beta_{1}}=&-\nabla_{a}hR^{am}R^{2}+h\nabla_{b}\left(R^{2}R^{m b}\right)+2R^{ab}R^{2}\nabla_{b}h^{m}_{a}-2h_{a}^{c}\nabla_{c}\left(R^{2}R^{am}\right)-\nabla^{m}h_{ab}R^{ab}R^{2}\nonumber\\
    &+h_{ab}\nabla^{m}\left(R^{ab}R^{2}\right)+2\nabla_{c}h^{cm}RR_{ab}R^{ab}-2h^{m d}\nabla_{d}\left(RR_{ab}R^{ab}\right)-2\nabla^{m}hRR_{ab}R^{ab}\nonumber\\
    &+2h\nabla^{m}\left(RR_{ab}R^{ab}\right),\\
    \Theta^{m}_{\beta_{2}}=&-\dfrac{3}{2}\nabla_{b}hRR^{ab}R_{a}^{m}+\dfrac{3}{2}h\nabla_{c}\left(RR^{am}R_{a}^{c}\right)+3\nabla_{c}h_{b}^{m}RR^{ab}R_{a}^{c}-3h_{b}^{d}\nabla_{d}\left(RR^{ab}R_{a}^{m}\right)\nonumber\\
    &-\dfrac{3}{2}\nabla^{m}h_{bc}RR^{ab}R_{a}^{c}+\dfrac{3}{2}h_{bc}\nabla_{m}\left(RR^{ab}R_{a}^{c}\right)+\nabla_{d}h^{dm}R_{a}^{c}R^{ab}R_{bc}-h^{em}\nabla_{e}\left(R_{a}^{c}R^{ab}R_{bc}\right)\nonumber\\
    &-\nabla^{m}h R_{a}^{c}R^{ab}R_{bc}+h\nabla^{m}\left(R_{a}^{c}R^{ab}R_{bc}\right),\\
     \Theta^{m}_{\beta_{3}}=&-2\nabla_{c}hR_{a}^{c}R^{ab}R_{b}^{m}+2h\nabla_{d}\left(R_{a}^{m}R^{ab}R_{b}^{d}\right)+4\nabla_{d}h_{c}^{m}R_{a}^{c}R^{ab}R_{b}^{d}-4h_{c}^{e}\nabla_{e}\left(R_{a}^{c}R^{ab}R_{b}^{m}\right)\nonumber\\
     &-2\nabla^{m}h_{cd}R_{a}^{c}R^{ab}R_{b}^{d}+2h_{cd}\nabla^{m}\left(R_{a}^{c}R^{ab}R_{b}^{d}\right),\\     
     \Theta^{m}_{\beta_{4}}=&-2\nabla_{c}hR_{ab}R^{ab}R^{cm}+
2h\nabla_{d}\left(R_{ab}R^{ab}R^{dm}\right)+4\nabla_{d}h_{c}^{m}R_{ab}R^{ab}R^{cd}-
4h_{c}^{e}\nabla_{e}\left(R_{ab}R^{ab}R^{cm}\right)\nonumber\\
&-2\nabla^{m}h_{cd}R_{ab}R^{ab}R^{cd}+2h_{cd}\nabla^{m}
\left(R_{ab}R^{ab}R^{cd}\right).
\end{align}

\paragraph{Noether-Wald charge:}
Having $\Theta$ in hand, by imposing the EoM the Noether-Wald $(d-2)$-form $Q_{\xi}$ can be read as
\begin{align}
    Q^{mn}_{\alpha_{0}}=&4\nabla^{[m}f^{\prime}\xi^{n]}-2f^{\prime}\nabla^{[m}\xi^{n]},\\
Q^{mn}_{\alpha_{1}}=&6\nabla^{b}\xi^{[m}R_{a}^{n]}R_{b}^{a}+6\xi^{[n}\nabla_{c}
\left(R^{m]a}R_{a}^{c}\right)+6\xi^{b}\nabla^{[m}\left(R^{n]a}R_{ab}\right),\\
Q^{mn}_{\alpha_{2}}=&4R\nabla^{a}\xi^{[m}R^{n]}_{a}+
2R^{ab}R_{ab}\nabla^{[n}\xi^{m]}+4\xi^{[n}\nabla_{b}\left(R^{m]b}R\right)+
4\xi^{[n}\nabla^{m]}\left(R_{ab}R^{ab}\right)\nonumber\\
&+4\xi^{b}\nabla^{[m}\left(R^{n]}_{b}R\right),\\
Q^{mn}_{\beta_{1}}=&4\xi^{[n}\nabla_{b}\left(R^{m]b}R^{2}\right)+4\nabla_{b}\xi^{[m}R^{n]b}R^{2}+4\xi_{b}\nabla^{[m}\left(R^{n]b}R^{2}\right)+4\nabla^{[n}\xi^{m]}RR_{ab}R^{ab}\nonumber\\
    &+8\xi^{[n}\nabla^{m]}\left(RR_{ab}R^{ab}\right)\\    Q^{mn}_{\beta_{2}}=&6\nabla_{b}\xi^{[m}R^{n]a}RR_{a}^{b}+6\xi^{[n}\nabla_{b}\left(R^{m]a}RR_{a}^{b}\right)+6\xi_{e}\nabla^{[m}\left(R^{n]a}RR_{a}^{e}\right)+2\nabla^{[n}\xi^{m]}R_{a}^{c}R^{ab}R_{bc}\nonumber\\
    &+4\xi^{[n}\nabla^{m]}\left(R_{a}^{c}R^{ab}R_{bc}\right),\\    Q^{mn}_{\beta_{3}}=&8\nabla_{d}\xi^{[m}R^{n]}_{a}R^{ab}R_{b}^{d}-8\xi^{[m}\nabla_{d}\left(R_{b}^{n]}R^{ab}R_{a}^{d}\right)+8\xi_{d}\nabla^{[m}\left(R^{n]}_{a}R^{ab}R_{b}^{d}\right),\\
    Q^{mn}_{\beta_{4}}=&8\nabla_{d}\xi^{[m}R^{n]d}R_{ab}R^{ab}+8\xi^{[n}\nabla_{d}
\left(R^{m]d}R_{ab}R^{ab}\right)+8\xi_{d}\nabla^{[m}\left(R^{n]d}R_{ab}R^{ab}\right).
\end{align}

\paragraph{Surface charges:}\label{surcha}
The covariant phase-space method yields the following expression for the surface charges associated with the diffeomorphisms, $\xi$,
\begin{equation}
    k^{mn}[\xi]=\delta Q^{mn}[\xi]-2\Theta^{[m}\xi^{n]}\, .
\end{equation}
Changing $Q_{\xi}$ with respect to the metric and using $\Theta$, one can find $k^{mn}$.
Finally, one can obtain the variation of the conserved charge associated with a given Killing vector $\xi$
\begin{equation}\label{eqcharge}
\delta H_{\xi}=\oint_{\partial \Sigma}k_{\xi}(\delta\phi,\phi).
\end{equation}
Since the surface charge expressions are very lengthy, we do not include them here. Now let us consider a stationary black hole solution with a Killing field $\xi$ which generates a Killing horizon and vanishes on a bifurcation surface $\mathcal{H}$. If we choose the hypersurface $\Sigma$ to have its outer boundary at spatial infinity and the interior boundary at $\mathcal{H}$, then the variational identity can be expressed with two boundary terms
\begin{equation}\label{eqtermo}
   \int_{H} k_{\xi} =\int_{\infty} k_{\xi},
\end{equation}
for $\delta_{\xi}\phi=0$. If we assume that the asymptotic symmetries have been specified by the time translational Killing field and axial rotational one with the horizon angular velocity $\Omega_{H}$, i.e., $\xi=\partial_{t}+\Omega_{H}\partial_{\phi}$. Then, the outer boundary integral of \eqref{eqtermo} can be defined as the total energy and the angular momentum.
Comparing \eqref{eqtermo} with the first law of thermodynamics $T\delta S=\delta M-\Omega_{H}\delta J$, the left-hand side gives the black hole entropy in the form
\begin{equation}\label{eqeqntropydef}
    S=\dfrac{2\pi}{\kappa}\int_{\mathcal{H}}Q_{\xi}.
\end{equation}
Here $\kappa$ is the surface gravity of the unperturbed black hole. The variation of the mass and angular momentum has the form
\begin{equation}\label{eqqmasang}
    \delta M=\dfrac{1}{16\pi G}\int_{\infty}k_{\xi}[\partial_{t}],\;\;\;\;\; \delta J=\dfrac{1}{16\pi G}\int_{\infty}k_{\xi}[\partial_{\phi}].
\end{equation}

\section{Metric Perturbations}\label{app3}
The metric of AdS$_{3}$ in light-cone coordinates $(x^{+},x^{-})$ on the boundary and a radial coordinate $\rho$ is given by
\begin{equation}
ds^{2}=-\dfrac{\ell^2}{4}\left[-4d\rho^2 +dx^{+2}+2\cosh(2\rho)dx^{+}dx^{-}+dx^{-2}\right].
\end{equation}
The AdS boundary is located at $\rho\to\infty$
\begin{equation}
    ds^{2}\sim \ell^2 d\rho^2-\dfrac{\ell^2}{4}e^{2\rho}dx^{+}dx^{-}+\mathcal{O}(1).
\end{equation}
Metric perturbations are written as
\begin{equation}\label{eqpert}
h_{\mu \nu}(\rho,x^{+},x^{-})=e^{-i(H x^{+}+Px^{-})}f_{\mu \nu}(\rho),
\end{equation}
where $f_{\mu\nu}(\rho)$ controls the radial fall-off. The components of radial part of $f_{\mu \nu}$ are given as follows
\begin{align}\label{eqpert1}
f_{++} &= \dfrac{1}{4}\cosh^{4-2H}\rho \tanh^{P-H}\rho (4\mathcal{C}_{2} \tanh^{2}\rho +\mathcal{C}_{1} \tanh^{4}\rho)\nonumber,\\
f_{+-} &= \dfrac{1}{2}\cosh^{2(1-H)}\rho \tanh^{P-H}\rho (\mathcal{C}_{2} \tanh^{2}\rho)\nonumber,\\
f_{+\rho} &= \dfrac{i}{32}\sinh^{-1}\rho \cosh^{-(1+2H)}\rho \tanh^{P-H}\rho (4(2\mathcal{C}_{2} -\mathcal{C}_{1})\cosh 2\rho -8\mathcal{C}_{2} +3\mathcal{C}_{1} +\mathcal{C}_{1} \cosh 4\rho)\nonumber\\
f_{--} &=0 \nonumber,\\
f_{-\rho} &=-\dfrac{i}{4}\cosh^{-1}\rho \sinh^{-1}\rho \sinh^{-H}2\rho \tanh^{P-H}\rho (\sinh^{H}2\rho \cosh^{-2H}\rho (-\mathcal{C}_{2} \cosh2\rho + \mathcal{C}_{2}))\nonumber,\\
f_{\rho \rho} &= \sinh^{-2-H}2\rho \tanh^{P-H}\rho (\cosh^{4-2H}\rho \sinh^{H}2\rho ((4\mathcal{C}_{2} -\mathcal{C}_{1})\tanh^{4}\rho)),
\end{align}
here, $f_{--}=0$ consistent with CSS boundary conditions, where $g_{--}$ is fixed.
The large $\rho$ behavior of the perturbed metric are given as
\begin{align}
f_{+\rho}&\sim \dfrac{i}{32}\left[4(2\mathcal{C}_{2}-\mathcal{C}_{1})e^{-2H\rho}+(3\mathcal{C}_{1}-8\mathcal{C}_{2})e^{-(2+2H)\rho}+\mathcal{C}_{1}e^{(2-2H)\rho}\right],\\
f_{-\rho}&\sim -\dfrac{i\mathcal{C}_{2}}{4}e^{-2(1+H)\rho}\left(1-e^{2H\rho}\right),\\
    f_{++} &\sim\left(\mathcal{C}_{1}+\dfrac{\mathcal{C}_{2}}{4}\right)e^{(4-2H)\rho},\\
    f_{+-}&\sim\; \dfrac{\mathcal{C}_{2}}{2}\;e^{(2-2H)\rho},\\
    f_{\rho\rho}&\sim \left(\mathcal{C}_{2}-\dfrac{\mathcal{C}_{1}}{4}\right)\;e^{-2H\rho} .
    \end{align}
Here, we have used  $\cosh(\rho)\propto e^{\rho},\sinh(\rho)\propto e^{\rho},\tanh(\rho)\propto 1$ at large $\rho$ $(\rho\gg 1)$. The CSS boundary conditions impose the following fall-offs on the metric perturbation
\begin{equation}
    g_{\mu\nu}=g_{\mu\nu}^{AdS}+h_{\mu\nu}
\end{equation}
with
\begin{equation}
    h_{+-}=\mathcal{O}(1),\;\;h_{++}=\mathcal{O}(1),\;\;\;h_{--}=\mathcal{O}(e^{-2\rho})
\end{equation}
The CSS boundary conditions do not constrain $P$, but require $H \geq 2$ to satisfy. Therefore, for $H=1$ and $0$, the CSS boundary conditions are violated.

\end{document}